\documentclass[12pt]{article}

\begin{document}

\author{$^{*}$T\^{a}nia Tom\'{e} and $^{+}$Alberto Petri \\
$^{*}$Instituto de F\'{\i}sica, Universidade de S\~{a}o Paulo \\
Caixa Postal 66318 \\
05315-970 S\~{a}o Paulo, SP, Brazil \\
$^{+}$Istituto di Acustica O. M. Corbino\\
Consiglio Nazionale delle Ricerche\\
Via del Fosso del Cavaliere, 100\\
00133 Roma, Italy}
\title{Cumulants of the three state Potts model and of nonequilibrium models
with C$_{3v}$ symmetry}
\date{}
\maketitle

\begin{abstract}
The critical behavior of two-dimensional stochastic lattice gas models with C%
$_{3v}$ symmetry is analyzed. We study the cumulants of the order parameter
for the three state (equilibrium) Potts model and for two irreversible
models whose dynamic rules are invariant under the symmetry operations of
the point group C$_{3v}$. By means of extensive numerical analysis of the
phase transition we show that irreversibility does not affect the critical
behavior of the systems. In particular we find that the Binder reduced
fourth order cumulant takes a universal value $U^{\ast }$ which is the same
for the three state Potts model and for the irreversible models. The same
universal behavior is observed for the reduced third-order cumulant.
\end{abstract}

\newpage\ 

\section{Introduction}

The critical behavior of nonequilibrium systems has been amply studied in
the last years \cite{dick,Priv,braz,kon}. These studies consider stochastic
lattice models and probabilistic cellular automata that evolves in time
according to an irreversible dynamics, that is, a dynamics that lacks
detailed balance. An important aspect to be considered is the role of
symmetry. Distinct systems with the same symmetry are expected to have
similar critical behavior. The symmetry is to be found in the Hamiltonian
for reversible systems and in the dynamics for the irreversible systems.
Among the irreversible models, there are models that have a reversible
counterpart with the same symmetries. In this context \ it has been
established the following statement: models with \textit{up-down} symmetry,
similar to the Ising model, and defined on the same lattice, reversible or
not, are in the same universality class \cite{grin}. This has been verified
numerically for a large number of models \cite%
{blote,tmm,mj,mendes,tamayo,nftd,wag}. We remark that the same universal
behavior is also observed for dynamic phase transitions in Ising models in
oscillating fields \cite{osc1,osc2,osc3}.

Recently \cite{bt1,14,bt2}, we have argued that this statement can be
extended to models with other symmetries. In fact, in these works, we
considered probabilistic cellular automata with dynamics that possess $%
C_{3v} $ symmetry and verified that the values of the critical exponents,
both static and dynamic, for irreversible systems are the same of the \emph{%
equilibrium} three-state Potts. That is, irreversibility is irrelevant
regarding the values of the critical exponents in systems with the symmetry
of the Potts models. In this article we complete this analyzes by performing
a systematic study of the cumulants at the critical point. We consider
stochastic lattice gas models with $C_{3v}$ symmetries: the three state
Potts model \cite{wu} and two irreversible models, and focus our attention
in the determination of the third and fourth order cumulants of the order
parameter. We present a Monte Carlo study of these properties for models
defined in a regular square lattice. Our results show that the value of the
cumulants, at the critical point, obtained for the present nonequilibrium
models, and the ones associated to the two-dimensional three-state Potts
(equilibrium) model are the same within numerical errors.

The paper is organized as follows. In Section 2 we present the models to be
studied. In Section 3, it is defined the quantities of interest in the study
of the phase transition and their scaling properties. The values of the
cumulants in the limit of infinite temperature and zero temperature are
discussed in Section 4. Sections 5 and 6 show numerical calculations and the
conclusions remarks.

\section{Models}

Consider a regular lattice of $N$ sites in which each site can be in one of
three states. At each site we attach a stochastic variable $\sigma _{i}$
that takes the values 1, 2 and 3. The state of the system can be represented
by $\sigma =(\sigma _{1},\sigma _{2},...,\sigma _{N})$. The time evolution
equation for $P(\sigma ,t)$, the probability of state $\sigma $ at time $t$,
is given by the master equation 
\begin{equation}
\frac{d}{dt}P(\sigma ,t)=\sum_{\sigma ^{\prime }}\{W(\sigma \mid \sigma
^{\prime })P(\sigma ^{\prime },t)-W(\sigma ^{\prime }\mid \sigma )P(\sigma
,t)\}  \label{mas}
\end{equation}
where the sum is over the $3^{N}$ configurations of the system. $W(\sigma
\mid \sigma ^{\prime })$ is the transition rate from a state $\sigma
^{\prime }$ to state $\sigma $, given that at the previous time step the
system was in state $\sigma ^{\prime }$. We will consider dynamics where $%
\sigma $ and $\sigma ^{\prime }$ can differ only by one site which we call a
one site dynamics. In this case we have

\begin{equation}
\frac{d}{dt}P(\sigma ,t)=\sum_{\alpha }\sum_{i=1}^{N}\{w_{i}(\sigma
^{i\alpha })P(\sigma ^{i\alpha },t)-w_{i}(\sigma )P(\sigma ,t)\}
\label{master}
\end{equation}%
where $w_{i}(\sigma )$ is the transition probability per site and 
\begin{equation}
\sigma ^{i\alpha }=(\sigma _{1},\sigma _{2},...,\sigma _{i}+\alpha
,\,...,\,\sigma _{N})  \label{aaa}
\end{equation}%
with $\alpha $ $=1,2.$ The states are defined modulo 3.

The transition probability $W(\sigma \mid \sigma ^{\prime })$ is invariant
under certain symmetry operations, that is, $W(R\sigma \mid R\sigma ^{\prime
})=W(\sigma \mid \sigma ^{\prime })$ where $R$ is a symmetry operation. For
the present models the symmetry operations $R$ are those that act on all
sites transforming each of them in the same manner. One of the symmetry
operations is the rotation operation $1\rightarrow 2$, $2\rightarrow 3$, and 
$3\rightarrow 1.$ Another is the operator $2\rightleftharpoons 3$ with state 
$1$ fixed. If the three states are placed on the vertices of a equilateral
triangle they correspond, respectively, to a rotation operation by 120
degrees and a specular operation. These symmetry operations define then the
point group C$_{3v}$.

\subsection{Equilibrium model and Metropolis prescription}

The Hamiltonian of the 3-state Potts model is 
\begin{equation}
H=\sum_{(ij)}-J\delta (\sigma _i,\sigma _j)  \label{hamil}
\end{equation}
where $\sigma _i=1,\,2,\,3$ , $J>0$ is the interacting strength between the
nearest neighbor sites $i$ and $j$ and $\delta $ is the Kronecker delta.

To simulate the model we associate to it a dynamics. We consider an one-site
dynamics as described by the master equation (\ref{master}). In the case the
model to be analyzed is an equilibrium model it is necessary to use a
transition probability $w_{i}(\sigma )$ that satisfy detailed balance
condition. That is, in the stationary state we must have 
\begin{equation}
w_{i}(\sigma ^{i\alpha })P(\sigma ^{i\alpha },t)=w_{i}(\sigma )P(\sigma ,t).
\label{detail}
\end{equation}
This dynamics can be defined by using the Metropolis prescription. We choose
a site $i$ and a state $\alpha $ and calculate 
\begin{equation}
w_{i}(\sigma )=\min \{1,\exp (-\beta \Delta H)\}  \label{metro}
\end{equation}
where $\Delta H=H(\sigma )-H(\sigma ^{i\alpha })$ is the difference between
the energy of the state $\sigma $ and the energy of \ the state $\sigma
^{i\alpha }$. The parameter $\beta $ is associated to the inverse of the
temperature $T$. Numerically we studied the critical point associated to the
model by performing Monte Carlo simulations. This procedure is described in
Section 6.

\subsection{Nonequilibrium models}

The nonequilibrium models are defined as follows. For the case of a square
lattice we are denoting the transition probability $w_0(\sigma )$ by $%
w(\sigma _0|\sigma _1,\sigma _2,\sigma _3,\sigma _4)$, where the sites 1, 2,
3, and 4 are the first neighbors of site 0.

\smallskip\ 

\textit{Symmetric stochastic lattice gas model}

\smallskip\ 

(a) If in the neighborhood of a given site there is a majority of sites
which are in one state then, independently of the state of the site, it
changes to the state of the majority with probability $p$. It changes to one
of the two other states with probability $(1-p)/2$.

(b) If no state is in majority then the site assume either state with equal
probability.

According to the local rules of the model we have 
\[
w(1\mid 1111)=w(1\mid 1112)=w(1\mid 1113)=w(1\mid 1123)=p 
\]%
and 
\begin{equation}
w(1\mid 1122)=w(1\mid 1133)=w(1\mid 2233)=1/3  \label{a3}
\end{equation}%
\smallskip The other rules are obtained by permutation of the neighboring
sites and by cyclic permutation of the states.

\smallskip\ 

\textit{Majority stochastic lattice gas model}

\smallskip \smallskip

The model consists of a stochastic lattice gas model where the site
transition probabilities follows similar rules of the majority vote model %
\cite{mj}. The chosen site adopts the same value of the majority of the
nearest neighbor sites with probability $p$. It adopts the state of the
other states with probability $q/2=(1-p)/2$. If there is an equal number of
nearest neighbors sites in the same state then the chosen site adopts each
state with probability $p/2$ or it assumes the other state with probability $%
q$. That is:

\[
w(1\mid 1111)=w(1\mid 1112)=w(1\mid 1113)=w(1\mid 1123)=p 
\]

\[
w(1\mid 1122)=w(1\mid 1133)=p/2 
\]
\[
w(1\mid 2233)=(1-p) 
\]
\begin{equation}
w(1\mid 1222)=w(1\mid 1333)=w(1\mid 3222)=w(1\mid 3312)=(1-p)/2  \label{a7}
\end{equation}
The other rules are obtained by permutation of the neighboring sites and by
cyclic permutation of the states.

It is straightforward to check that the transition probabilities $%
w_{i}(\sigma )$, for both models, are invariant under the symmetry
operations of the group $C_{3v}$.

The nonequilibrium models, have the same symmetries as the Hamiltonian of
the three-state Potts model, given in (\ref{hamil}) although in the present
case the models are not defined by a Hamiltonian and do not satisfy detailed
balance condition (\ref{detail}). That is, these models are microscopically
irreversible.

\section{Cumulants and scaling properties}

A convenient way to analyze the present models is through the use of the
variables 
\begin{equation}
x_\alpha =\frac 1N\sum_{i=1}^N(\delta (\sigma _i,\alpha )-\frac 13)
\label{ss1}
\end{equation}
where $\alpha $ assumes the values $1,\,2$ and $3$ and $\delta (x,y)$ is the
Kronecker delta. The order parameter has three components $x_1,\,x_2,\,$ and 
$x_3$ but just two of them are independent and the following property 
\[
x_1+x_2+x_3=0, 
\]
holds.

It is useful also to introduce a set of homogeneous functions $%
I_n(x_1,x_2,x_3)$, of a given order $n$, that are invariant under the
symmetry operations $R$ defined above. There is just one independent second
order invariant given by 
\begin{equation}
I_2=\frac 13\left( x_1^2+x_2^2+x_3^2\right) ,  \label{ss2}
\end{equation}
and just one independent third order invariant 
\begin{equation}
I_3=-\frac 29\left( x_1^3+x_2^3+x_3^3\right) ,  \label{ss3}
\end{equation}

The fourth order invariant function is 
\begin{equation}
I_{4}=\frac{2}{9}\left( x_{1}^{4}+x_{2}^{4}+x_{3}^{4}\right) .  \label{i4}
\end{equation}
Again there is just one independent fourth order invariant.

\subsection{Cumulants}

In the present study the quantities of interest are the order parameter, 
\begin{equation}
m=\left| \overrightarrow{m}\right| =\left\langle \sqrt{I_{2}}\right\rangle ,
\label{order}
\end{equation}
and the reduced cumulant, 
\begin{equation}
U_{24}=\frac{\left\langle I_{4}\right\rangle }{\left\langle
I_{2}\right\rangle ^{2}}  \label{u24}
\end{equation}
and, also, the Binder fourth order cumulant, which in terms of $I_{4}$ and $%
I_{2}$ above defined is given by 
\begin{equation}
U=1-\frac{1}{3}U_{24}.  \label{binder}
\end{equation}

We also analyzed the behavior of the order three reduced cumulant,

\begin{equation}
U_{23}=\frac{\left\langle I_{3}\right\rangle }{\left\langle
I_{2}\right\rangle ^{3/2}},  \label{u23}
\end{equation}

\subsection{Scaling properties}

The order parameter $\overrightarrow{m}$ has two independent components $x$
and $y$ and we will denote 
\begin{equation}
\overrightarrow{m}=\frac{1}{\sqrt{2}}(x\overrightarrow{i}+y\overrightarrow{j}%
)  \label{vetor}
\end{equation}
where $x$ and $y$ are related to $x_{1}$, $x_{2}$ and $x_{3}$ by the
relations 
\begin{eqnarray}
x_{1} &=&-\frac{\sqrt{3}}{2}x-\frac{1}{2}y,  \label{x1x} \\
x_{2} &=&\frac{\sqrt{3}}{2}x-\frac{1}{2}y,  \nonumber \\
x_{3} &=&y.
\end{eqnarray}

The invariants (\ref{ss2}), (\ref{ss3}) and (\ref{i4}) are written, in terms
of $x$ and $y$, as 
\begin{equation}
I_2=\frac 12(x^2+y^2)  \label{ii2}
\end{equation}
\begin{equation}
I_3=\frac 12(x^2y-\frac{y^3}3)  \label{ii3}
\end{equation}
\begin{equation}
I_4=\frac 14(x^4+2x^2y^2+y^4)  \label{ii4}
\end{equation}

Our main interest is to calculate the moments of the distribution associated
to the order parameter. The moment $M_{n}$of order $n$ can be
defined by 
\[
M_{n}=<|m|^{n}>=\int |m|^{n}P(m,\epsilon ,L)dm. 
\]%
where $P(m,\epsilon ,L)$ is the probability distribution of $m=|%
\overrightarrow{m}|,$ where $\overrightarrow{m}=\frac{1}{\sqrt{2}}(x,y)$,
and $\epsilon $ is equal to the deviation of the external parameter from its
critical value and $L$ is the system size. We assume that 
\begin{equation}
P(m,\epsilon ,L)=L^{\beta /\nu }\phi (m/\epsilon ^{\beta },L\epsilon ^{\nu
}).  \label{scal1}
\end{equation}%
Defining $z=m/\varepsilon ^{\beta }$, 
\[
M_{n,L}=\epsilon ^{\beta (n+1)}L^{\beta /\nu }\int |z|^{n}\phi (z,L\epsilon
^{\nu })dz. 
\]%
Then, 
\begin{equation}
M_{n,L}=\epsilon ^{\beta n}(\epsilon ^{\nu }L)^{\beta /\nu }F_{n}(\epsilon
L^{1/\nu })  \label{escal2}
\end{equation}%
where $F_{n}(Y)$ is an universal function.

From relation (\ref{escal2}), we get the following scaling forms 
\begin{equation}
M_{L}=<m>=L^{-\beta /\nu }\widetilde{m}(\epsilon L^{1/\nu })  \label{scalm}
\end{equation}
\begin{equation}
\frac{M_{L}^{4}}{(M_{L}^{2})^{2}}=\widetilde{U}_{24}(\epsilon L^{1/\nu }),
\label{escal4}
\end{equation}
and 
\begin{equation}
\frac{M_{L}^{3}}{(M_{L}^{2})^{3/2}}=\widetilde{U}_{23}(\epsilon L^{1/\nu }).
\label{escal3}
\end{equation}
\qquad where $\epsilon $ is the deviation of the external parameter from its
critical value and $\widetilde{m}(x)$, $\widetilde{U}_{23}(x)$ and $%
\widetilde{U}_{24}(x)$ are universal functions.

For an infinite system the correlation length diverges as $\xi \sim \epsilon
^{-\nu }$ and the scaling forms give the behavior $m\sim \epsilon ^{\beta }$
for the order parameter. Moreover, the reduced cumulants $U_{24}$ and $%
U_{23} $, defined in equations (\ref{u24}) and (\ref{u23}), are expected to
attain according to (\ref{escal4}) and (\ref{escal3}) a universal value at
the critical point, that does not depend on the lattice size. The same
behavior, of course, must hold for the reduced fourth-order cumulant $U$
given in equation (\ref{binder}).

\section{Exact results}

When the temperature $T$ $\rightarrow 0$, the equilibrium model defined by
the Hamiltonian (\ref{hamil}), will be in the ordered phase and with
probability $1$ in one of the three Potts states. For the nonequilibrium
models, defined in (\ref{a3}) and (\ref{a7}), this limits correspond to $%
p\rightarrow 1$.

In this limit it is expected the following behavior 
\begin{equation}
<(x_{\alpha })^{n}>\rightarrow \{(\frac{1}{3})(\frac{2}{3})^{n}+(\frac{1}{3}%
)(-\frac{1}{3})^{n}+(\frac{1}{3})(-\frac{1}{3})^{n}\}.  \label{x1n}
\end{equation}
with $\alpha =1,\,2,\,$ and $3$ and $n=1,\,2,\,3,....$. Depending on the
initial conditions the system will be with $\sigma _{i}=1$, for all sites in
the lattice, or in state where $\sigma _{i}=2$, for all sites in the
lattice, or in the state $\sigma _{i}=3$, for all sites in the lattice. The
factor $(\frac{1}{3})$ in equation (\ref{x1n}) takes into account this fact.
For example, the second order invariant defined in expression (\ref{ss2})
will attain, the value 
\[
I_{2}\rightarrow \frac{1}{3}\{(\frac{1}{3})(\frac{2}{3})^{2}+(\frac{1}{3})(-%
\frac{1}{3})^{2}+(\frac{1}{3})(-\frac{1}{3})^{2}\} 
\]
that is $I_{2}\rightarrow 2/9\,$ when\thinspace $T\rightarrow
0\,\,(p\rightarrow 1).$

Following the same procedure, the limiting values of the\ third and fourth
order invariants and reduced cumulants can be easily evaluated. In
particular the values of the fourth order cumulant, defined in (\ref{u24}),
in the limit $T\rightarrow 0$\thinspace $(p\rightarrow 1)$, will be 
\begin{equation}
U_{24}\rightarrow 1.  \label{lim24}
\end{equation}%
which implies that the Binder fourth order cumulant, according to equation (%
\ref{binder}), takes the limit, 
\begin{equation}
U\rightarrow \frac{2}{3}.  \label{limu}
\end{equation}

On the other hand when the temperature $T$ $\rightarrow \infty $ $%
\,(p\rightarrow 1/3)$ the equilibrium system (the nonequilibrium symmetric
model), is in the disordered state, 
\[
x_{1}=x_{2}=x_{3}=0. 
\]%
So the probability distribution associated to the order parameter is a
distribution of independent variables, a Gaussian distribution, and we can
write 
\begin{equation}
P(x,y)=\frac{1}{2\pi a}\exp [-(x^{2}+y^{2})/2a],  \label{Gauss}
\end{equation}%
with 
\[
a=<(x^{2}+y^{2})/2>. 
\]%
So in this limit we have that the expressions (\ref{ii2}) and (\ref{ii4})
are related by 
\begin{equation}
I_{4}=3(I_{2})^{2}.  \label{gauss24}
\end{equation}%
Which implies that 
\begin{equation}
U_{24}\rightarrow 3  \label{li24}
\end{equation}%
and 
\begin{equation}
U\rightarrow 0.  \label{li25}
\end{equation}

The third order invariant $I_{3}\rightarrow 0$ when $T\rightarrow \infty
\,\,\,(p\rightarrow 1/3)$. So $U_{23}\rightarrow 0.$

\section{Monte Carlo simulations}

The system evolves in time according to the local rules and eventually
reaches a steady state that can be of two types: a disordered steady state,
where there is an equal average number of sites in each one of the three
Potts states; either an ordered steady state characterized by the
predominance of sites in one of the Potts states.

The simulation of the equilibrium and the nonequilibrium models with $C_{3v}$
symmetry was performed by considering square lattices with $L^{2}=N$ sites,
and periodic boundary conditions. Each simulation started with a
configuration generated at random and averages over several simulations
where taken to get the final results.

\subsection{Equilibrium model}

We consider several values of the external parameter, the temperature $T.$
We pick a site $i$ at random and then we apply the Metropolis prescription
to update site $i$ according to the expression (\ref{metro}) as follows. Let
the state of site $i$ be $\sigma _{i}$. We change the site variable to $%
\sigma _{i}+\alpha $ and calculate $\Delta H\,$according to expression (\ref%
{hamil}) considering the nearest neighbor sites of the site $i$ (which have
not changed, since we are considering a one site dynamics). If $\Delta H\leq
0$, then the new state will be $\sigma ^{i\alpha }=(\sigma _{1},...,\sigma
_{i+\alpha },...,\,\sigma _{N})$. Otherwise, if $\Delta H>0,$we calculate $%
p=\exp (-\beta \Delta H)$ and generate a random number $\zeta $ equally
distributed in the interval $[0,1]$. If $\zeta \leq p$ then the new state
will be $\sigma ^{i}$, otherwise the state does not change, that is, remains 
$\sigma $.

The system evolves in time until it reaches a stationary state. The time
taken to the system to reach the stationary state depends on the temperature
and on the lattice size. After discarding the first configurations, we used
the following states in order to evaluate the state functions cumulants of
the distribution probability associated to the order parameter. The
stationary states are equilibrium stationary states, i.e. they satisfy the
detailed balance condition. As expected we found two types of stationary
states: a ordered one where the order parameter $m$ is different from zero
and a disordered one where $m=0$.

The critical temperature for two dimensional three state Potts model is
given by $k_{B}T_{c}=1/(\ln (\sqrt{3}+1)\simeq 0.99497$ \cite{wu}. In our
simulation we take $k_{B}=1$ and we analyze the behavior of the cumulants (%
\ref{u24}) and (\ref{u23}) as a function of the temperature and for
different lattice sizes. As we can see in Figure 1 and Figure 2 when $%
T\rightarrow T_{c}$ the cumulant $U_{24}\rightarrow U_{24}^{\ast }$ and the
cumulant $U_{23}$ $\rightarrow U_{23}^{\ast }$ , respectively, and it
follows that $U_{24}^{\ast }$ and $U_{23}^{\ast }$ are universal. That is in
the critical point $U_{24}$ and $U_{23}$ attain universal values that do not
depend on the lattice size. We used the finite size scaling relations (\ref%
{escal3}) and (\ref{escal4}). The expression (\ref{escal3}) is related to
the reduced Binder \cite{bin} fourth order cumulant that so assumes an
universal value $U^{\ast }$. The value for these functions at the critical
point where found by us to take 
\begin{equation}
U_{24}^{\ast }=1.16\pm 0.01 \label{uu4}
\end{equation}%
and $U^{\ast }\widetilde{=}0.61$. The value of $U_{23}$ at the critical
point (see Figure 2) is $U_{23}^{\ast }=0.245\pm 0.01.$

\subsection{Nonequilibrium models}

We consider several value of the parameter $p$. At each time step just one
site is chosen at random and it is updated according to the prescriptions
given in Section 2.2 are applied (rules (\ref{a3}) for the symmetric model
and rules (\ref{a7}) for the majority model). After a transient, which
depends on the model, on the size of the system and on the value of $p$, the
system attains a steady state. Our simulations show that both models exhibit
continuous phase transitions with the ordered steady state ($m\neq 0$)
occurring at high values of $p$. As $p$ is decreased the transition takes
place at a critical value $p_{c}$, which is different for each model, and
the system becomes disordered ($m=0$) for $p$ less than $p_{c}$.

Using the finite size scaling relations (\ref{escal3}) and (\ref{escal4}) we
obtain the critical value $p_{c}$ for each model. For the \emph{symmetric
lattice gas model},\ as shown in Figure 3 and Figure 4, the curves of $%
U_{24} $ versus $p$ and the curves of $U_{23}$ versus $p$, for different
values of $L$, intercept at the critical point $p_{c}$ estimated to be $%
p_{c}=0.892\pm 0.003$. It is worth to call the attention that this model is
similar to the one considered by one of us in a previous work \cite{14}.
Both models evolves in time according to the same local Markovian rules.
However the one considered in reference \cite{14} is a probabilistic
cellular automaton (synchronous update) whereas the present model evolves in
time according to a sequential dynamics (asynchronous update). As
irreversible models are defined by the dynamics itself we do not have to
expect the same value of the critical parameter. In fact they are different.
We also observe that the results obtained previously \cite{14} for the
fourth order cumulant for the probabilistic cellular automaton are not
sufficient precise due to presence of large fluctuations. Contrastingly, in
the study of the stochastic lattice gas models considered here, we verified
that the behavior of the cumulants, both of third and fourth order, are
smooth.

Figure 5 and Figure 6 show the curves of $U_{24}$ versus $p$ and the curves
of $U_{23}$ versus $p$ for different values of $L$ for the \emph{majority}
stochastic lattice gas model. The interception of these curves yields the
critical point estimated to be $p_{c}=0.883\pm 0.001$. The values attained
by $U_{23}$ and $U_{24}$ are universal and are $U_{23}^{\ast }=0.244\pm 0.01$
and $U_{24}^{\ast }=1.16\pm 0.01.$ And the Binder reduced fourth order
cumulant is $U^{\ast }\widetilde{=}0.61.$ These universal values are in
agreement with the results for the (equilibrium) three state Potts model.

\section{Summary}

We have considered systems that undergo a phase transition from a state with
high symmetry to a state with lower symmetry. The phase with high symmetry
is invariant under the symmetry operations of the symmetry group $C_{3v}$.
We analyzed first, the equilibrium three states Potts model, and then two
irreversible models. All models have a continuous time evolution, governed
by a master equation. In the first case the model is defined by an
Hamiltonian whereas in the second case they are defined only by transitions
rates that do not obey the detailed balance condition. The phase transition
that takes place, as an external parameter is varied, is a continuous phase
transition from a disordered steady state to an ordered steady state. For
the case of the equilibrium model, the phase transition occurs when the
temperature $T$ is varied and for the nonequilibrium cases when the
parameter $p$ is varied. We introduced a set of homogeneous functions that
are invariant under the symmetry operations of the group C$_{3v}$. From
these functions we define the order parameter and the cumulants. The
critical points were estimated by numerical simulations on regular square
lattices of different sizes and using finite size scaling theory. Analyzing
the cumulants we conclude that irreversibility plays an irrelevant role in
the critical behavior and is not a property that might change the universal
behavior. In fact, the fourth-order and the third-order cumulants attains
universal values at the critical point and we found that this values are the
same for the equilibrium system and for nonequilibrium systems, whenever
periodic boundary conditions are considered. It is worthwhile to observe
that the value of the cumulants may depend on the boundary conditions as was
established for the equilibrium two dimensional models \cite{derrida}.

\bigskip

\textbf{Acknowledgments}

This work was partially supported by the Fundac\~{a}o de Amparo \`{a}
Pesquisa do Estado de S\~{a}o Paulo (FAPESP) through the 01-09590-8 project.

\begin{center}
\textbf{Figure Captions}
\end{center}

Figure 1.The reduced cumulant $U_{24}$ as a function of $T$ for $L=14,\,20,$ 
$28,$ and $40$ (square lattices) for the three state (equilibrium) Potts
model.

\smallskip\ 

Figure 2.The reduced cumulant $U_{23}$ as a function of $T$ for $L=14,$ $%
20,\,$\ $28$ and $40$ (square lattices) for the three state (equilibrium)
Potts model.

\smallskip\ 

Figure 3.The reduced cumulant $U_{24}$ as a function of $p$ for $L=14,$ $%
20,\,$\ $28$ and $40$ (square lattices) for the three state symmetric model
(nonequilibrium).

\smallskip\ 

Figure 4.The reduced cumulant $U_{23}$ as a function of $p$ for $L=14,$ $%
20,\,$\ $28$ and $40$ (square lattices) for the symmetric model
(nonequilibrium).

\smallskip\ 

Figure 5.The reduced cumulant $U_{24}$ as a function of $p$ for $L=14,$ $%
20,\,$\ $28$ and $40$ (square lattices) for the three state (nonequilibrium)
majority model.

\smallskip\ 

Figure 6.The reduced cumulant $U_{23}$ as a function of $p$ for $L=14,$ $%
20,\,$\ $28$ and $40$ (square lattices) for the three state (nonequilibrium)
majority model.

\end{document}